\documentclass[aps, pra, preprint, amssymb, groupedaddress, showpacs]{revtex4-1}

\usepackage{amsfonts}
\usepackage{amsmath}
\usepackage{amssymb}
\usepackage{microtype}
\usepackage{graphicx}
\usepackage{hyperref}
\usepackage[utf8]{inputenc}
\usepackage[T1]{fontenc}
\usepackage[usenames,dvipsnames]{xcolor}

\newcommand{\bra}[1]{\langle #1|}
\newcommand{\ket}[1]{|#1\rangle}
\newcommand{\braket}[2]{\langle #1|#2\rangle}

\begin{document}

\title{
Role of two-electron processes in the excitation-ionization of lithium
atoms by fast ion impact}

\author{T. Kirchner}
\email[]{tomk@yorku.ca}

\author{N. Khazai}

\affiliation{Department of Physics and Astronomy, York University, Toronto, Ontario, Canada, M3J 1P3}

\author{L. Guly\'{a}s}
\affiliation{Institute for Nuclear Research, Hungarian Academy of Sciences (ATOMKI),
P.O. Box 51, H-4001 Debrecen, Hungary}

\date{\today}

\begin{abstract}
We study excitation and ionization in the 1.5 MeV/amu O$^{8+}$-Li
collision system, which was the subject of a recent reaction-microscope-type
experiment [Fischer \textit{et al.}, Phys. Rev. Lett. \textbf{109}, 113202 (2012)].
Starting from an independent-electron model based on determinantal wave functions 
and using single-electron basis generator method and
continuum distorted-wave with eikonal initial-state
calculations we show that pure single ionization
of a lithium $K$-shell electron 
is too weak a process to explain the
measured single differential cross section.
Rather, our analysis suggests that
two-electron excitation-ionization processes occur
and have to be taken into account when comparing with the data. 
Good agreement is obtained only if 
we replace the
independent-electron calculation by an independent-event model
for one of the excitation-ionization processes and also take
a shake-off process into account.
\end{abstract}


\maketitle

\section{Introduction}
\label{sec:intro}
The investigation and differentiation of single and multiple electron processes 
has been a topic of considerable interest in atomic collision physics
over many years.
For one thing, studies in this area shed light on 
questions relevant for applied research, e.g., in
radiation therapy, where the damage induced by swift
ions is different in single- and multiple-ionization events~\cite{gervais06}.
Multiple ionization of, e.g., water molecules results in fragmentation 
practically with certainty, while there is a high chance that the
H$_2$O$^+$ ion created after single-electron removal stays intact~\cite{murakami12}.

Much of the experimental and theoretical activity in the investigation of single
and multiple processes is, however, 
fueled by the fundamental interest in the few-body quantum
dynamics at play.
The seemingly simplest situation that can occur in an ion-atom collision
corresponds to single-electron removal, in which one target electron is either captured
by the (bare) projectile or promoted to a continuum state. 
However, single-electron removal is not necessarily a pure one-electron process: Another 
target electron may participate in the dynamics and end up in an excited state 
after the collision. In the case of capture, these so-called transfer plus target-excitation (TTE)
events have been identified in cold target recoil ion momentum
spectroscopy (COLTRIMS) and reaction microscope (ReMi) experiments, which give access to the
$Q$-value of a given reaction, i.e., the electronic energy loss or gain~\cite{hasan06,schoeffler09,guo12}.
Excitation-ionization (EI) was measured some time ago at very high projectile
energy using (Auger-) electron spectroscopy~\cite{tanis99}.

If the target atom is helium, which has often been the case in COLTRIMS 
and ReMi experiments,
TTE and EI are true two-electron processes.
A somewhat different situation arises if more than one target shell is occupied
before the collision.
In this case, the single capture or single ionization of an inner-shell electron leaves the
target ion behind in an excited state. In a naive independent-particle picture 
a two-electron process,
in which an outer-shell electron is removed and an inner-shell electron is promoted to the
state just vacated, has the same outcome.
This type of two-electron TTE or EI in principle becomes distinguishable 
from pure single inner-shell electron removal
if the inner-shell electron is promoted to a target state that was vacant
before the collision\footnote{It is indistinguishable though from another two-electron
process that involves inner-shell electron removal
and outer-shell electron excitation; see the discussion of Eq.~(\ref{eq:1svac-result}) 
in Sec.~\ref{sec:theory}}. 
In practice, however, the resolution achievable in $Q$-value measurements may make it difficult to
separate these two-electron processes from pure one-electron removal. 
Theoretical calculations are then required for
a full understanding of the situation.

A recent joint experimental-theoretical work on
1.5 MeV/amu O$^{8+}$collisions from lithium atoms was concerned with this problem~\cite{fischer12}. 
The experiments were performed with the
newly-developed 'MOTReMi' apparatus, which combines a magneto-optical trap (MOT) to cool the lithium
atoms with a ReMi to measure the reaction products.
The recorded $Q$-value spectra for single ionization exhibit two distinct peaks, which can be
associated with the ejection of the $2s$ valence electron and the ejection of an inner $K$-shell
electron, respectively. 
A continuum distorted-wave with
eikonal initial-state (CDW-EIS) calculation for Li($2s$) ionization was found to be in reasonable
agreement with the measured
electron-energy single differential cross section (SDCS) corresponding to the first peak, 
but a CDW-EIS calculation 
for Li($1s$) ionization differed markedly from the data in the other channel.
It was concluded that
two-electron EI, which was not taken into account in the calculation,
is needed to explain the measurement.

In this paper, we provide a theoretical analysis
of the O$^{8+}$-Li collision system based on the 
independent-electron (IEL) model to scrutinize this
interpretation.
The IEL model is discussed in Sec.~\ref{sec:theory}.
In Sec.~\ref{sec:results}, we present
IEL SDCSs for various pure ionization and EI processes and compare them with
the experimental data. Extensions of the IEL model, among them an independent-event
(IEV) model for one EI process, are considered to account for
the quite substantial discrepancies.
A summarizing discussion is provided in Sec.~\ref{sec:summary}.
Atomic units characterized by $\hbar=m_e=e=4\pi\varepsilon_0=1$ are used unless otherwise stated.

\section{Independent-electron treatment of pure single ionization and excitation-ionization}
\label{sec:theory}

Within the semiclassical approximation (SCA) the
O$^{8+}$-Li collision system is described by a time-dependent Schr\"odinger equation (TDSE)
for the electronic Hamiltonian
\begin{equation}
           \hat H_e(t)  = \sum_{i=1}^{N} [\hat T_i + \hat V_i^{en}(t)] + \sum_{i<j}^N \hat W_{ij} ,
\label{eq:he}
\end{equation}
which consists of single-electron kinetic energy operators $\hat T_i$, 
electron-nucleus interactions $\hat V_i^{en}$ (which depend on time due to
the SCA assumption of classically moving nuclei),
and the electron-electron Coulomb interactions $\hat W_{ij}$.
To the best of our knowledge an explicit solution of this correlated $N$=3-electron problem
has not been attempted yet.

The IEL model consists in replacing the Hamiltonian (\ref{eq:he}) by a one-body operator
\begin{equation}
           \hat H_e(t) \rightarrow \sum_{i=1}^{N} \hat{h}_i(t) 
\label{eq:iem}
\end{equation}
such that the TDSE separates into a set of single-particle equations for the three
electrons. 
We assume the
single-particle Hamiltonian to be of the form
\begin{equation}
   \hat h(t) = -\frac{1}{2}\Delta + V_{\rm Li}(|\mathbf{r}_t|)- \frac{Z_P}{|\mathbf{r}_p|},
\label{eq:h1}
\end{equation}
where 
$Z_P$ is the charge number of the bare projectile ion, and
$\mathbf{r}_t$ and $\mathbf{r}_p$ are the position vectors of the electron with respect to the target
and the projectile center, respectively. They are related according to
$\mathbf{r}_p = \mathbf{r}_t-\mathbf{R}(t)$ with $\mathbf{R}(t)$ being the classical straight-line
trajectory of the projectile relative to the target center.
We note that the Laplace operator in Eq.~(\ref{eq:h1}) is taken with respect to
the center-of-mass reference frame.
The effective potential
$V_{\rm Li}$ represents the interactions in the 
($1s^2 2s$) ground-state configuration of the lithium atom. It is obtained from
the exchange-only version of the optimized potential method (OPM) of 
density functional theory~\cite{opm},
i.e., it includes electron-nucleus Coulomb interactions, screening, and exchange terms exactly 
and exhibits the correct asymptotic $-1/r_t$ behavior, but it neglects electron
correlations.

\begin{figure}[ht]
\begin{center}
\includegraphics[width = 0.65\linewidth]{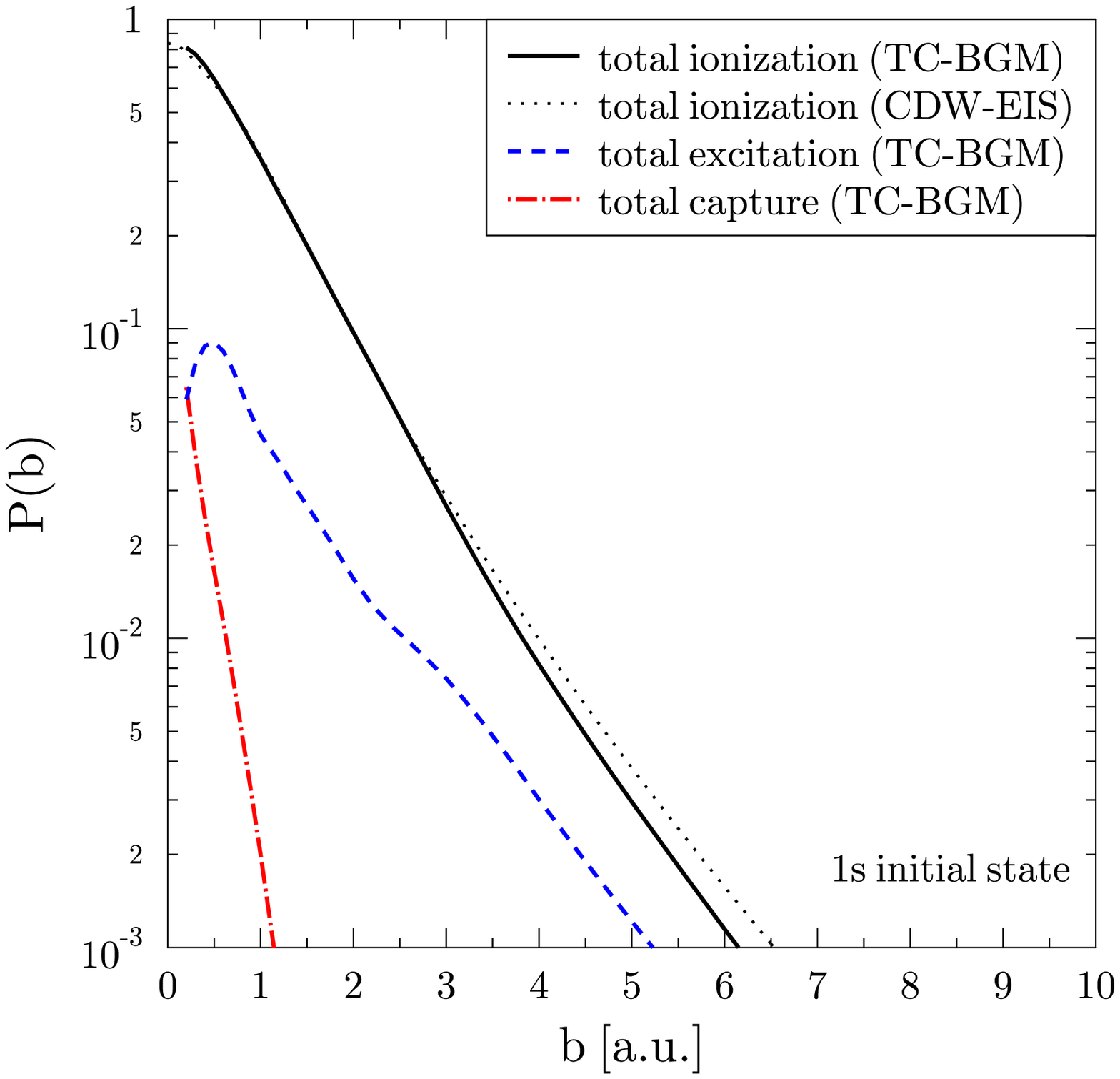}

\vspace{-2\baselineskip}
\includegraphics[width = 0.65\linewidth]{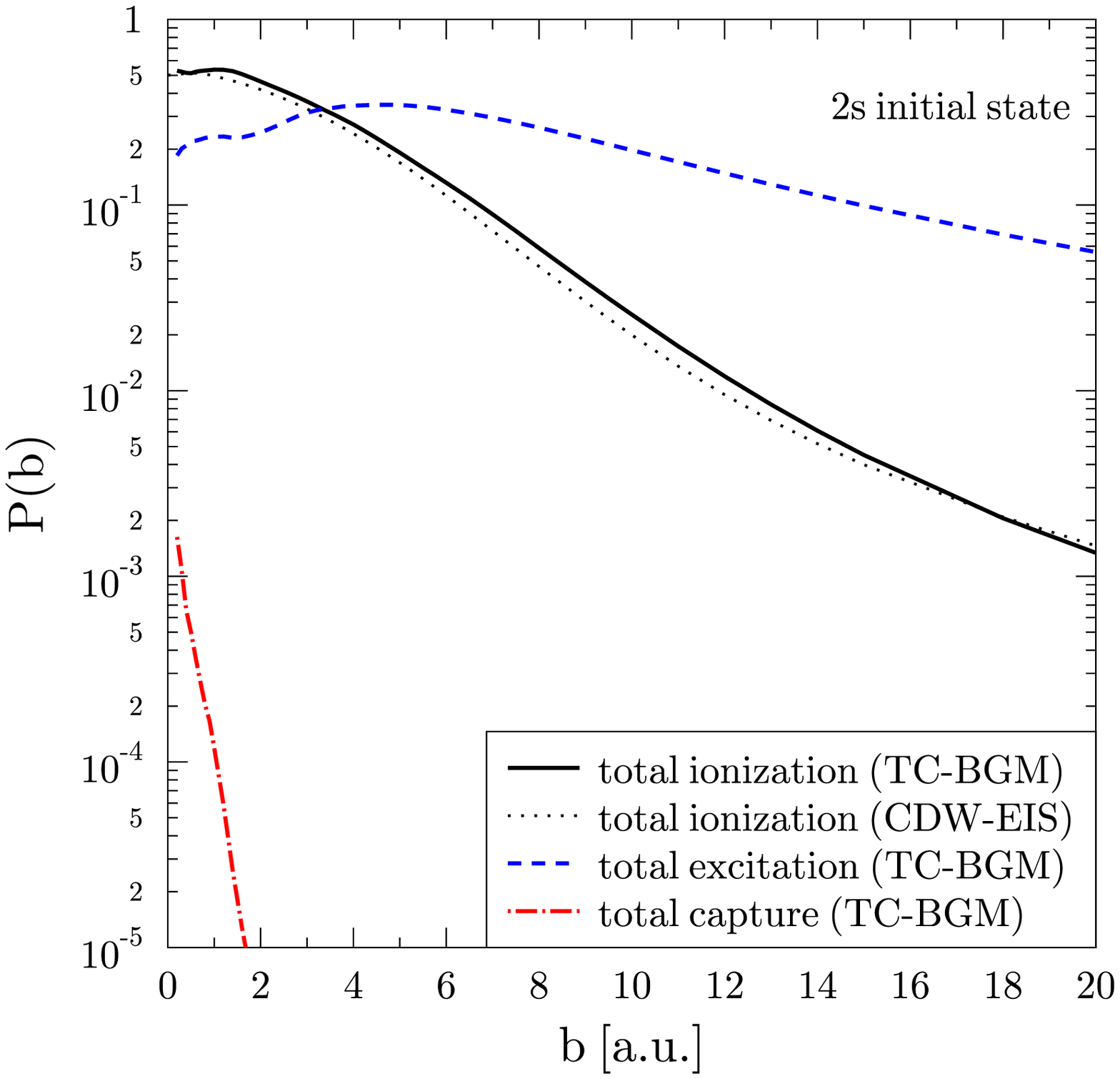}
\caption{(Color online) Total excitation, capture, and ionization 
probabilities according to Eqs.~(\ref{eq:p1sexc}) to (\ref{eq:p2sion}) 
for 1.5 MeV/amu O$^{8+}$-Li collisions obtained from TC-BGM calculations
for Li($1s$) (upper panel) and Li($2s$) (lower panel) initial states. CDW-EIS ionization
probablilities are displayed in addition as (black) dotted curves.
\label{fig:single}}
\end{center}
\end{figure}

We have solved the single-particle equations for the Hamiltonian (\ref{eq:h1}) 
and the initially occupied Li($1s$) and Li($2s$) orbitals using the two-center basis generator
method (TC-BGM) with a basis that consists of the $1s-4f$ target states, the $1s-4f$ 
(hydrogenlike) projectile
states, as well as 71 BGM pseudo states to account for ionization~\cite{tcbgm}.
All basis states are endowed with electron translation factors to ensure Galilean invariance.
Results for the total excitation ($p_{1s(2s)}^{\rm exc}$),
capture ($p_{1s(2s)}^{\rm cap}$) and ionization ($p_{1s(2s)}^{\rm ion}$) probabilities
are shown in Fig.~\ref{fig:single} as functions of the impact parameter $b$.
The probabilities are calculated as follows:

\begin{eqnarray}
    p_{1s\rightarrow f_k} & = & |\braket{f_k}{\psi_{1s}(t_f)}|^2 
\label{eq:p1s-fk} \\
    p_{2s\rightarrow f_k} & = & |\braket{f_k}{\psi_{2s}(t_f)}|^2 
\label{eq:p2s-fk} \\
    p_{1s}^{\rm exc}   & = &    \sum_{f_k\in T, f_k \neq 1s} p_{1s\rightarrow f_k} 
\label{eq:p1sexc} \\
    p_{2s}^{\rm exc}   & = &    \sum_{f_k\in T, f_k\neq \{1s,2s\}} p_{2s\rightarrow f_k} 
\label{eq:p2sexc} \\
    p_{1s}^{\rm cap} & = &    \sum_{f_k\in P} p_{1s\rightarrow f_k} 
\label{eq:p1scap} \\
    p_{2s}^{\rm cap} & = &    \sum_{f_k\in P} p_{2s\rightarrow f_k} 
\label{eq:p2scap} \\
    p_{1s}^{\rm ion}   & = &    
            1 - \sum_{f_k\in T} p_{1s\rightarrow f_k} -  \sum_{f_k\in P} p_{1s\rightarrow f_k} 
\label{eq:p1sion} \\
    p_{2s}^{\rm ion}   & = &    
            1 - \sum_{f_k \in T} p_{2s\rightarrow f_k} -  \sum_{f_k \in P} p_{2s\rightarrow f_k} ,
\label{eq:p2sion}
\end{eqnarray}
where $\ket{\psi_{1s(2s)}(t_f)}$ denote the solutions of the
single-particle equations corresponding to the $1s (2s)$ initial states 
and propagated to a sufficiently large final time $t_f$ after
the collision, and $\ket{f_k} \in T(P)$ are 
final target (projectile) states. 
Note that in Eqs.~(\ref{eq:p1sion}) and (\ref{eq:p2sion}) 
we use the unitarity of the TC-BGM solutions of the single-particle equations.

As expected at a projectile energy as high as 1.5 MeV/amu, 
capture is very weak except for the $1s$ initial state in close collisions.
It will be neglected in the following, i.e., we will identify electron removal from the lithium
atom with ionization into the continuum.
We further observe in Fig.~\ref{fig:single} that ionization strongly dominates 
for the case of the $1s$ initial state, while
for the Li($2s$) initial state excitation takes over in distant
collisions. The details of our calculations show that excitation to $2p$ is the
strongest channel.
We also include in Fig.~\ref{fig:single} ionization probabilities 
obtained from a CDW-EIS calculation
for the Hamiltonian (\ref{eq:h1}) \cite{crothers83,fainstein96}. 
They are in very good agreement with the
TC-BGM results for the case of the $1s$ initial state, and still in acceptable agreement
for $2s$---keeping in mind that the CDW-EIS method is perturbative in nature and the
perturbation parameter $\eta=Z_P/v$  
is close to one ($\eta = 1.03 $ for the projectile speed $v=7.75$ a.u.).

In order to make contact with the experiment \cite{fischer12} and 
take EI processes into account, we have to reinstate many-body aspects of the
collision system. Consistent with the IEL model this is done by assembling 
the solutions of the single-particle equations 
in the form of a Slater determinant.
If the three-electron final states are also taken to be Slater determinants, the 
transition probabilities of interest can be obtained without further approximation 
from combinations of determinants constructed from one-particle density matrix elements
\cite{hjl85}
\begin{equation}
\bra{f_k}\hat{\gamma}^1(t_f)\ket{f_l}
 =  \sum_{i=1}^N \braket{f_k}{\psi_i(t_f)} \braket{\psi_i(t_f)}{f_l}  .
\label{eq:gamma1}
\end{equation}
In this work, we are interested in processes, in which
exactly one vacancy is created in the lithium atom. They correspond to
probabilities for finding two bound single-particle target states, say
$\ket{f_1}$ and $\ket{f_2}$, occupied
and all the others vacant, and can be
calculated according to \cite{hjl85}
\begin{equation}
    P_{f_1 f_2}^{\sum_k \bar{f}_k}  \equiv  P_{f_1 f_2} - 
                  \sum_{f_k\in T} P_{f_1 f_2 f_k} .
\label{eq:parthole}
\end{equation}
In the expression on the right hand side of Eq.~(\ref{eq:parthole})
$P_{f_1 f_2}$ denotes the \textit{inclusive} probability for finding two electrons
in the subconfiguration
$\ket{f_1 f_2}$ while nothing is known about the final state of the third electron.
As shown in Ref.~\cite{hjl85} this inclusive probability is given as the 
determinant constructed from the $2\times 2$ density matrix corresponding to 
$\ket{f_1 f_2}$.
$P_{f_1 f_2 f_k}$ is the
\textit{exclusive} probability to find the three electrons in the completely specified configuration 
$\ket{f_1 f_2 f_k}$ and is given as the determinant of the $3\times 3$ density matrix 
corresponding to $\ket{f_1 f_2 f_k}$.
Since the sum in Eq.~(\ref{eq:parthole}) runs over all bound target states the difference
of both terms 
corresponds to the statement that one of the electrons has been removed from the lithium atom.

For the explicit evaluation of Eq.~(\ref{eq:parthole}) we have to take the spin projections
($\uparrow , \downarrow$) of the electrons into account and consider spin orbitals
that we denote, e.g., by writing $\ket{f_k\uparrow}$.
Since the Hamltonian (\ref{eq:h1}) is spin independent we have 
$\braket{f_k\uparrow}{\psi_i\uparrow}=\braket{f_k\downarrow}{\psi_i\downarrow}=\braket{f_k}{\psi_i}$
and
$\braket{f_k\uparrow}{\psi_i\downarrow}=\braket{f_k\downarrow}{\psi_i\uparrow}=0$.
We choose the initial state of the lithium atom to be a Slater determinant
built from the 
$(1s\hskip-3pt\uparrow 1s\hskip-3pt\downarrow 2s\hskip-3pt\uparrow)$ spin orbitals.

With these preparations we are ready to
consider the probabilities that correspond to the experimentally
distinguishable processes \cite{fischer12}:

(i) $2s$ vacancy production $P_{2s}^{\, \rm vac}$, in which 
the Li$^+$ ion is found in its ($1s^2$) ground state:
\begin{equation}
   P_{2s}^{\, \rm vac} =  P_{1s\uparrow 1s\downarrow}^{\sum_k \bar{f}_k\uparrow} ,
\label{eq:2svac}
\end{equation}

(ii) $1s$ vacancy production $P_{1s}^{\, \rm vac}$, in which the Li$^+$ ion is left
in the excited configuration ($1snl$) with $n\ge 2$:
\begin{equation}
   P_{1s}^{\, \rm vac} = \sum_{f_l\in T,f_l \neq 1s}
                         (P_{1s\uparrow f_l\uparrow}^{\sum_k \bar{f}_k\downarrow}
                         + P_{1s\uparrow f_l\downarrow}^{\sum_k \bar{f}_k\uparrow} 
                         + P_{1s\downarrow f_l\uparrow}^{\sum_k \bar{f}_k\uparrow}) .
\label{eq:1svac}
\end{equation}

All terms on the right hand sides of Eqs.~(\ref{eq:2svac}) 
and (\ref{eq:1svac}) can be computed using the prescription (\ref{eq:parthole}).
It is instructive to work out the determinants analytically.
With the definitions (\ref{eq:p1s-fk}) to (\ref{eq:p2sion}) and
\begin{eqnarray}
    p_{1s}^{\rm elast} & = &    p_{1s\rightarrow 1s} 
\label{eq:p1selast} \\
    p_{2s}^{\rm elast} & = &    p_{2s\rightarrow 2s} 
\label{eq:p2selast}
\end{eqnarray}
one obtains 
\begin{eqnarray}
   P_{2s}^{\, \rm vac} & = & P_{2s}^{\, \rm excl} +
      P_{2s}^{\, \rm ex}  + \Delta P_{2s}^{\, \rm anti} 
\label{eq:2svac-result} \\
   P_{1s}^{\, \rm vac} & = &  P_{1s}^{\, \rm excl} +
       P_{EI1} + P_{EI2} +  P_{1s}^{\, \rm ex} + \Delta P_{1s}^{\, \rm anti} 
\label{eq:1svac-result}
\end{eqnarray}
with
\begin{eqnarray}
   P_{2s}^{\, \rm excl} & = & (p_{1s}^{\rm elast})^2  p_{2s}^{\rm ion} 
\label{eq:2sexcl} \\
   P_{2s}^{\, \rm ex}   & = & p_{1s}^{\rm elast}  p_{1s}^{\rm ion} p_{2s\rightarrow 1s} 
\label{eq:2sex} \\
   P_{1s}^{\, \rm excl} & = & 2p_{1s}^{\rm elast} p_{1s}^{\rm ion} p_{2s}^{\rm elast} 
\label{eq:1sexcl} \\
   P_{EI1} & = & 2 p_{1s}^{\rm elast} p_{1s}^{\rm ion} p_{2s}^{\rm exc} 
\label{eq:ei1} \\
   P_{EI2} & = & 2 p_{1s}^{\rm elast} p_{1s}^{\rm exc} p_{2s}^{\rm ion} 
\label{eq:ei2} \\
   P_{1s}^{\, \rm ex}  & = & 2 p_{1s}^{\rm exc}  p_{1s}^{\rm ion} p_{2s\rightarrow 1s} 
\label{eq:1sex}
\end{eqnarray}
and correction terms 
$\Delta P_{1s}^{\, \rm anti}$ and $\Delta P_{2s}^{\, \rm anti}$
stemming from off-diagonal elements of the density matrix that reflect the antisymmetry of
the three-electron wave functions. These terms as well as
a few details regarding the derivation of Eqs.~(\ref{eq:2svac-result}) 
and (\ref{eq:1svac-result}) are given in the Appendix.

Equations (\ref{eq:2sexcl}) to (\ref{eq:1sex}) have straightforward interpretations and can
also be obtained from a simple multinomial analysis of the three-electron problem that
is based on associating each electron with a single-particle
probability. Expressions~(\ref{eq:2sexcl}) and (\ref{eq:1sexcl}) correspond to
exclusive ionization events, in which the non-ionized electrons remain bound in their
initial states. The probabilities~(\ref{eq:ei1}) and (\ref{eq:ei2}) correspond to
two-electron EI processes, in which one $K$-shell electron does not change its state,
while the other two electrons are either excited or ionized.
Finally, Eqs.~(\ref{eq:2sex}) and (\ref{eq:1sex}) describe exchange processes
which arise as a consequence of the indistinguishability of the electrons.
In practice, these exchange terms
as well as the correction terms $\Delta P_{1s}^{\, \rm anti}$ and $\Delta P_{2s}^{\, \rm anti}$
are negligible. 

\begin{figure}[ht]
\begin{center}
\includegraphics[width = \linewidth]{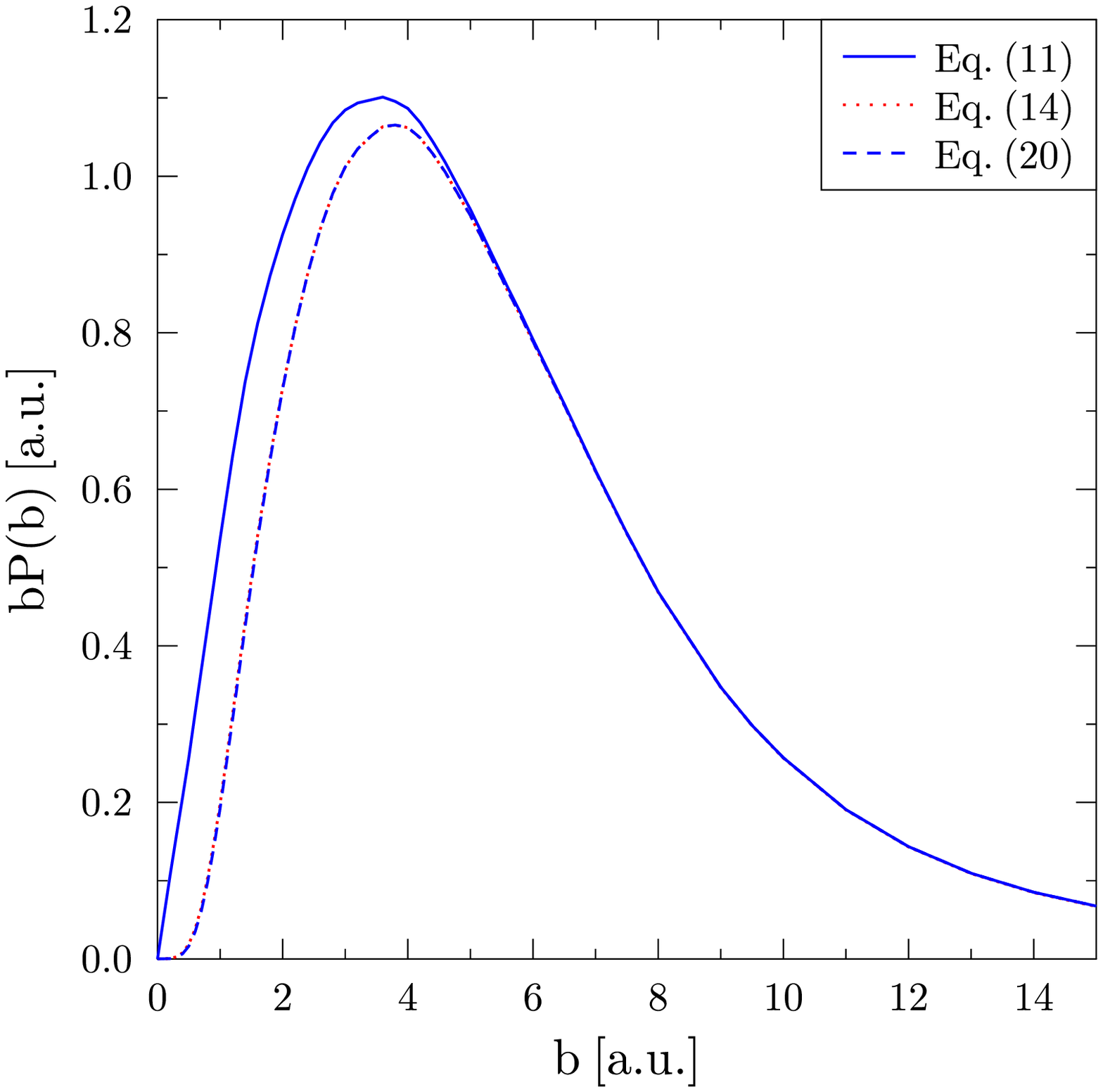}
\caption{(Color online) Impact-parameter-weighted 
probabilities for $2s$ vacancy production [Eq.~(\ref{eq:2svac})], 
exclusive $2s$ ionization [Eq.~(\ref{eq:2sexcl})], and single-particle $2s$ ionization 
[Eq.~(\ref{eq:p2sion})]
for 1.5 MeV/amu O$^{8+}$-Li collisions.
All curves shown are obtained from TC-BGM calculations.
The curves displaying the probabilities~(\ref{eq:2svac}) and~(\ref{eq:2sexcl}) sit on top of each other.
\label{fig:pb2s}}
\end{center}
\end{figure}

\begin{figure}[ht]
\begin{center}
\includegraphics[width = \linewidth]{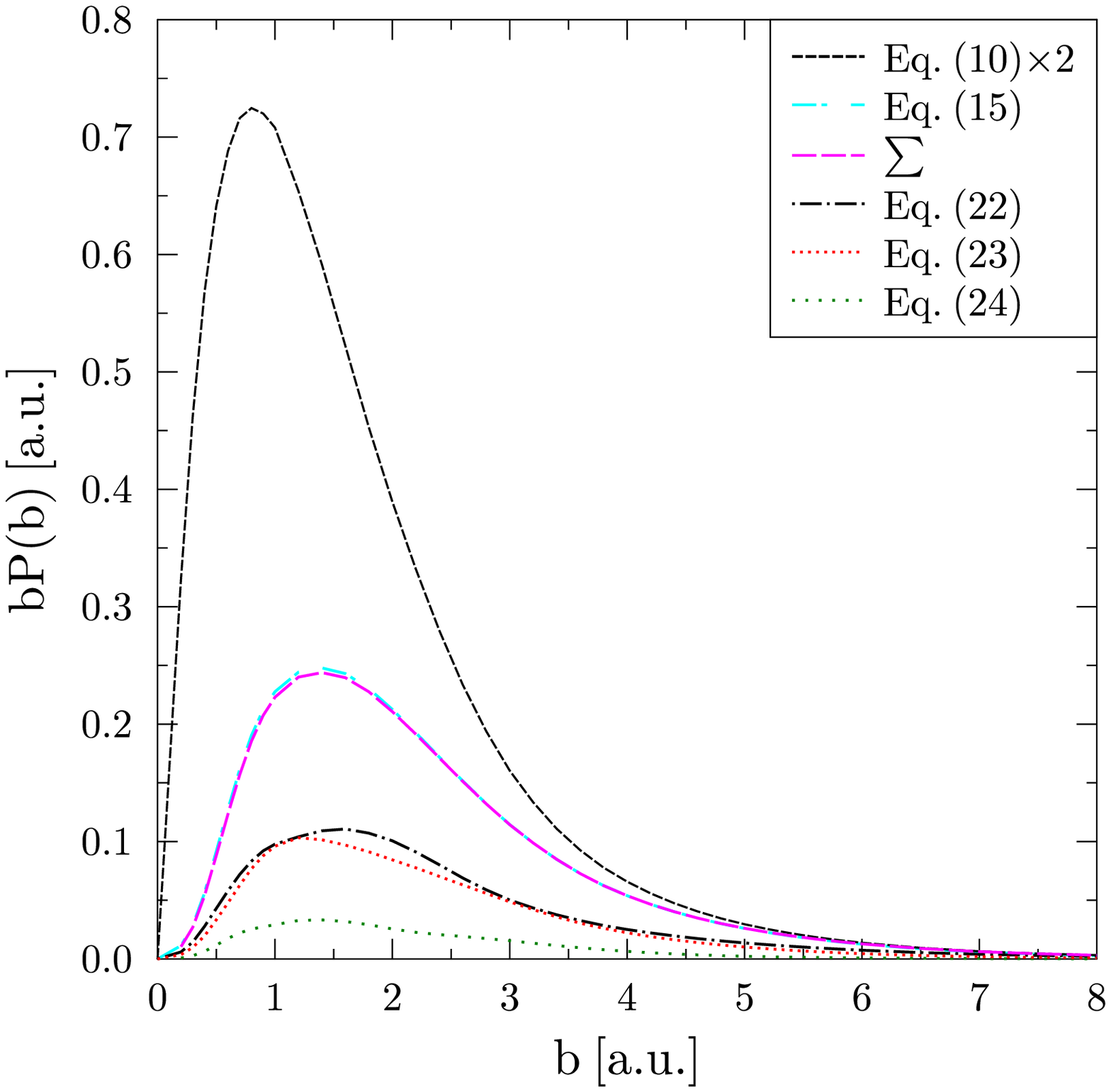}
\caption{(Color online) Impact-parameter-weighted probabilities for
$1s$ vacancy production [Eq.~(\ref{eq:1svac})],
exclusive $1s$ ionization [Eq.~(\ref{eq:1sexcl})],
$1s$ ionization with $2s$ excitation [Eq.~(\ref{eq:ei1})],
$1s$ excitation with $2s$ ionization [Eq.~(\ref{eq:ei2})],
the sum of (\ref{eq:1sexcl}), (\ref{eq:ei1}) and (\ref{eq:ei2})
($\sum$),
and single-particle $1s$ ionization multiplied by two
[Eq.~(\ref{eq:p1sion})$\times 2$]
for 1.5 MeV/amu O$^{8+}$-Li collisions.
All curves shown are obtained from TC-BGM calculations.
\label{fig:pb1s}}
\end{center}
\end{figure}

This is demonstrated in Figs.~\ref{fig:pb2s} and \ref{fig:pb1s}.
Figure~\ref{fig:pb2s} displays the $b$-weighted probability for $2s$ vacancy production
according to Eq.~(\ref{eq:2svac}). 
Differences to 
exclusive $2s$ ionization according to Eq.~(\ref{eq:2sexcl}) are too small to be visible
mainly because the exchange amplitude for a transition from the $2s$ to the $1s$ state 
in the correction terms is very small.
We have also included $p_{2s}^{\rm ion}$ in Fig.~\ref{fig:pb2s}. 
This probability is interpreted as the single-ionization probability in a one-active-electron (OAE)
model in which the two $K$-shell electrons are assumed to be frozen throughout the collision. 
Obviously, the OAE result differs from the other curves only at relatively small $b$, where the 
$1s$ electrons do undergo transitions with non-negligible probabilities.

In Fig.~\ref{fig:pb1s} we show $b$-weighted probabilities for processes that
result in the creation of one $K$-shell vacancy. 
$P_{EI1}$ [Eq.~(\ref{eq:ei1})] is almost as strong as $P_{1s}^{\,\rm excl}$ [Eq.~(\ref{eq:1sexcl})],
whereas $P_{EI2}$ [Eq.~(\ref{eq:ei2})], which involves the excitation of
a $K$-shell electron, is considerably weaker. The sum of these three 
probabilities is in almost perfect agreement with the
full result for $1s$ vacancy production according to
Eq.~(\ref{eq:1svac})---demonstrating that also for $1s$ ionization
the exchange and antisymmetry 
correction terms in Eq.~(\ref{eq:1svac-result}) 
can be neglected. 

In many calculations for helium targets the quantity $2p_{1s}^{\rm ion}$ was used to calculate ionization
cross sections (see, e.g., Ref.~\cite{gulyas08,gulyas12}). 
This procedure again corresponds to an OAE 
model, in which the factor of two arises since one does not know which of the two $K$-shell
electrons is active and which is passive.
We have included a $2p_{1s}^{\rm ion}$ curve in Fig.~\ref{fig:pb1s}. It is 
seen to be much larger than
the other probabilities, which indicates that the assumption of just one active $K$-shell
electron is not realistic for the 1.5 MeV O$^{8+}$-Li collision system. The main
reason for this is that 
it is very unlikely for the initial $2s$ electron
not to undergo a transition in a relatively close collision.
Put another way, if the assumption of two passive electrons is unrealistic, 
the quantity $2p_{1s}^{\rm ion}$ is contaminated by multielectron
processes.

\section{Differential cross section results and extensions of the IEL model}
\label{sec:results}

Having analyzed the relative strengths of various ionization processes within
the IEL model we now turn to
the SDCSs for the two reaction channels discussed in Ref.~\cite{fischer12}.
Ideally, we would calculate them on the basis of Eqs.~(\ref{eq:2svac}) and (\ref{eq:1svac})
and the TC-BGM single-particle solutions. However, this is not possible, 
since it is difficult to extract
electron-emission-energy ($E_e$) differential information from the population of the BGM pseudo states.
Instead, we use the CDW-EIS method to calculate single-particle ionization 
probabilities $p_i^{\rm ion}(b,E_e)$
for $i=1s,2s$ and combine them with the $b$-dependent TC-BGM probabilities for excitations and
elastic transitions according to the simplified Eqs.~(\ref{eq:2sexcl}), (\ref{eq:1sexcl}),
(\ref{eq:ei1}), and (\ref{eq:ei2}). 
The resulting $b$ and $E_e$ dependent probabilities
are integrated over the impact parameter in order
to obtain the $E_e$-differential SDCSs.
Given that (i) the exchange and antisymmetry correction terms in Eqs.~(\ref{eq:2svac-result}) 
and (\ref{eq:1svac-result})
were shown to be small, and (ii) the CDW-EIS total ionization probabilities $p_i^{\rm ion}(b)$
were shown to be close to their TC-BGM counterparts this procedure is unlikely to
produce errors in addition to those stemming from the limitations of the IEL model.

\begin{figure}[ht]
\begin{center}
\includegraphics[width = \linewidth]{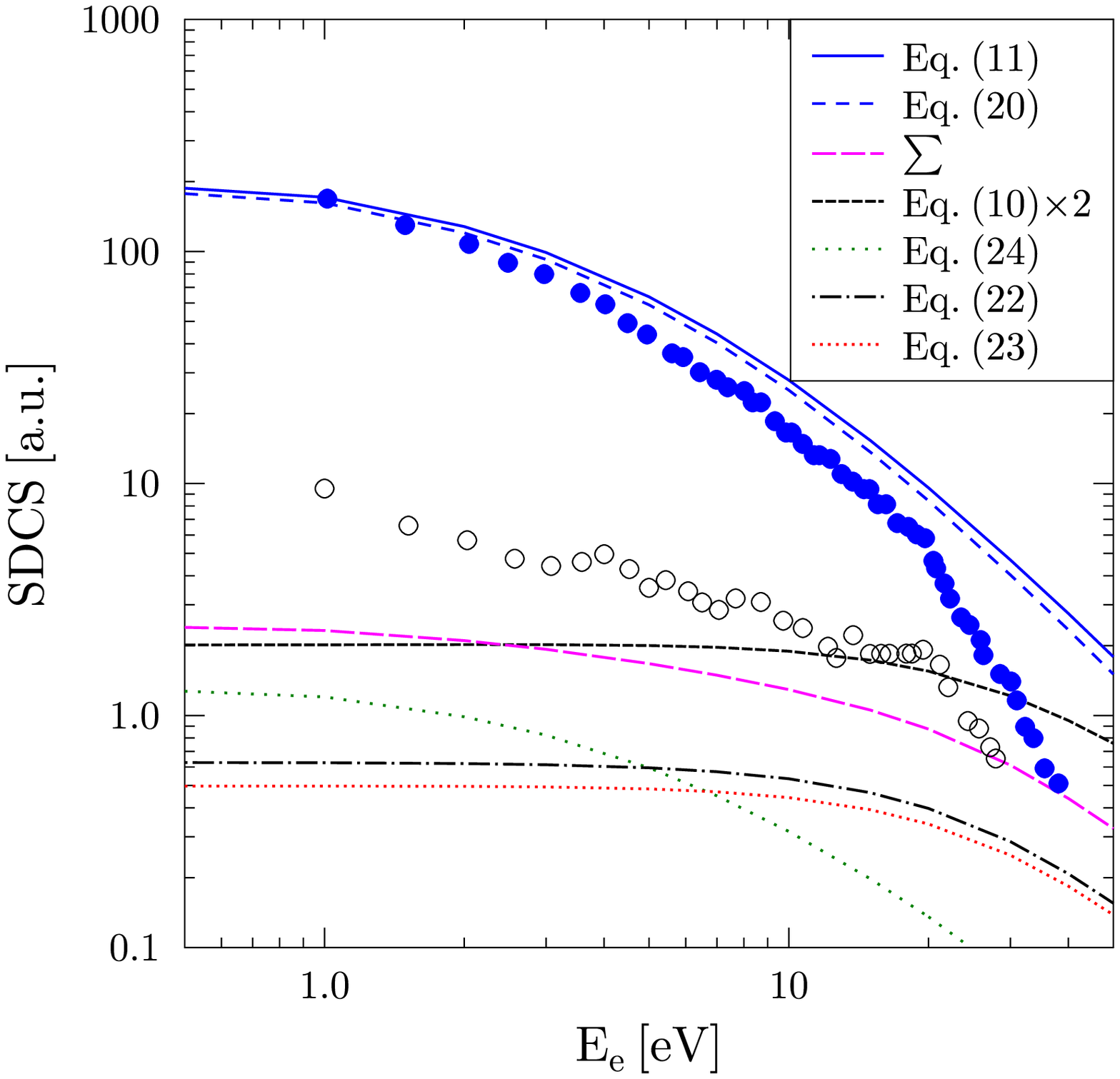}
\caption{(Color online) Single differential cross sections (SDCSs) for
single-particle $2s$ ionization [Eq.~(\ref{eq:p2sion})],
exclusive $2s$ ionization [Eq.~(\ref{eq:2sexcl})], 
single-particle $1s$ ionization multiplied by two [Eq.~(\ref{eq:p1sion})$\times 2$],
exclusive $1s$ ionization [Eq.~(\ref{eq:1sexcl})],
$1s$ ionization with $2s$ excitation [Eq.~(\ref{eq:ei1})],
$1s$ excitation with $2s$ ionization [Eq.~(\ref{eq:ei2})], and
the sum of (\ref{eq:1sexcl}), (\ref{eq:ei1}) and (\ref{eq:ei2})
($\sum$)
as functions of the electron energy
for 1.5 MeV/amu O$^{8+}$-Li collisions.
Experimental data: \cite{fischer12}. 
\label{fig:sdcs1}}
\end{center}
\end{figure}

Results are shown in Fig.~\ref{fig:sdcs1} in comparison with the experimental
data of Ref.~\cite{fischer12}.
For the $2s$ channel the OAE and exclusive ionization SDCSs are very similar 
(cf.~Fig.~\ref{fig:pb2s}) and in reasonable agreement with the data, which were deemed 
low at energies $E_e>20$ eV due to acceptance limitations of the
spectrometer~\cite{fischer12}. We note that the present results are in very good agreement
with the CDW-EIS calculations shown in Fig.~2(b) of Ref.~\cite{fischer12}, which we did not
include in our figure for the sake of clarity.

For the $1s$ channel the situation is more involved. Our OAE result agrees
well with the CDW-EIS calculation reported in Ref.~\cite{fischer12} (not shown), 
but it differs in both magnitude and shape from the experimental data.
As mentioned earlier, the OAE cross section is contaminated by multielectron processes,
particularly by multiple ionization, which
is excluded in the measured coincidences of electrons and singly-charged ions.
Hence, this quantity is not very useful for the analysis of the experimental data.
On the other hand, exclusive $1s$ ionization is much smaller 
than the measurements at all electron energies. 
The same is true for the
EI process described by Eq.~(\ref{eq:ei1}), which
consists of $1s$ ionization and $2s$ excitation.
Similarly to exclusive $1s$ ionization 
this EI process reflects the energy dependence of the OAE $1s$ ionization curve, 
which can be inferred from Eqs.~(\ref{eq:1sexcl}) and (\ref{eq:ei1}). 

Likewise, the other EI process [Eq.~(\ref{eq:ei2})]
which involves $2s$ ionization and $1s$ excitation reflects the energy dependence of the
$2s$ ionization curve. If one adds it to exclusive $1s$ ionization and
$1s$ ionization with $2s$ excitation to calculate the total IEL SDCS for $1s$-vacancy
production one obtains a curve whose shape has some similarity with the energy dependence of the
experimental data, but lies significantly below them except at the highest electron energies.
We thus have to conclude that the IEL model is not sufficient to explain the measurements.

In an attempt to understand the discrepancy without embarking on a full solution of
the correlated three-electron problem we have considered the following scenarios.
First, one may argue that for
the EI process corresponding to Eq.~(\ref{eq:ei2}) 
a sequential IEV picture of the collision dynamics is more realistic than the IEL model. 
In such a scenario the outer ($2s$) electron is
removed first with the result that the inner ($1s$) electrons rearrange before
one of them is promoted to an excited state
of the Li$^+$ ion.
Accordingly, we also
considered a Hamiltonian in which the OPM potential $V_{\rm Li}$ is replaced by the
OPM potential $V_{\rm Li^+}$ for the ($1s^2$) configuration which
decays like $-2/r_t$ asymptotically. The $1s$-excitation with $2s$-ionization SDCS
in the IEV model was then calculated according to
\begin{equation}
   \left(\frac{d\sigma}{dE_e}\right)_{\rm IEV} =  2 \int d^2 b \, 
      p_{1s+}^{\rm elast}(b) p_{1s+}^{\rm exc}(b) 
      p_{2s}^{\rm ion}(b,E_e) ,
\label{eq:iev}
\end{equation}
where  $p_{1s+}^{\rm elast}(b)$ and $p_{1s+}^{\rm exc}(b)$ denote single-particle
elastic and excitation probabilities obtained from TC-BGM collision calculations for O$^{8+}$-Li$^+$, 
and $p_{2s}^{\rm ion}(b,E_e)$ is the electron-energy differential CDW-EIS ionization probability
for O$^{8+}$-Li that we also used in the IEL calculations.
We note that we did not consider the IEV counterpart of the EI process (\ref{eq:ei1}), in which
a $K$-shell electron is removed and the $2s$ electron excited, since it does not seem plausible
to assume that the wave function of the outer $2s$ electron adjusts to an ionic ($1s2s$) configuration
after inner-shell ionization. Besides, we know from the analysis presented above that
such an IEV probability would result in the same $E_e$-dependence as 
OAE $1s$ ionization, which is at odds with the measured data.

\begin{figure}[ht]
\begin{center}
\includegraphics[width = \linewidth]{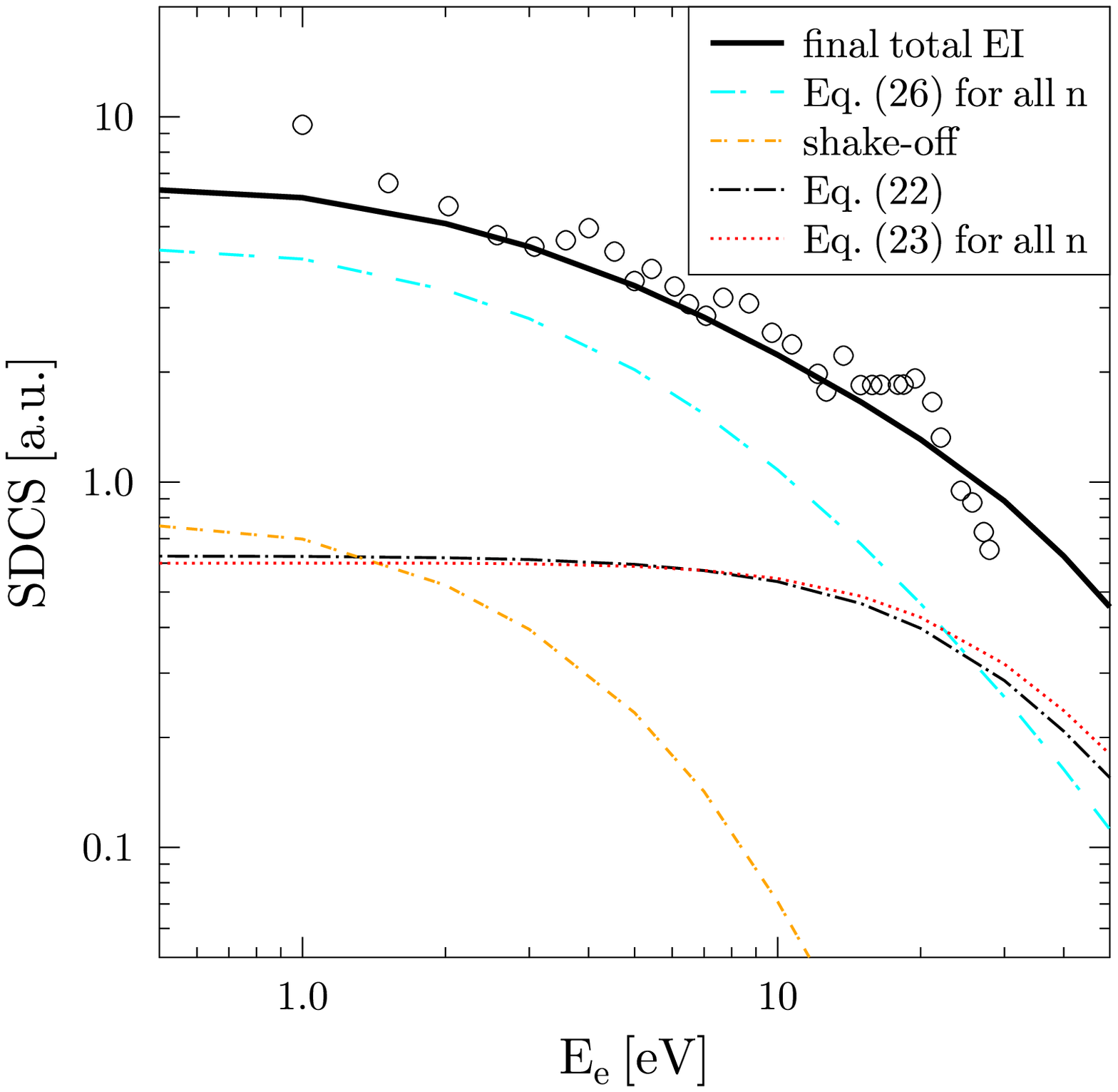}
\caption{(Color online) Single differential cross sections (SDCSs) for 
$2s$ ionization with $1s+$ excitation extrapolated to $n\rightarrow \infty$  [Eq.~(\ref{eq:iev}) for all $n$], 
the shake-off model as explained in the text,
exclusive $1s$ ionization [Eq.~(\ref{eq:1sexcl})], and
$1s$ ionization with $2s$ excitation extrapolated to $n\rightarrow \infty$ [Eq.~(\ref{eq:ei1}) for all $n$]
as functions of the electron energy for 1.5 MeV/amu O$^{8+}$-Li collisions. 
The full thick curve dubbed 'final total EI'
is obtained from adding the four other calculated SDCSs shown here and represents 
our final result for total EI.
Experimental data: \cite{fischer12}.
\label{fig:sdcs2}}
\end{center}
\end{figure}

Secondly, our TC-BGM basis only includes target states 
up to principal quantum number $n=4$. 
It is impossible to push the calculation further without
severely restricting the representation of the continuum. In order to estimate the contribution
from higher excitations we have assumed that they scale like $1/n^3$ which allows for
extrapolating them to $n\rightarrow\infty$~\cite{roehrbein10}. 

Finally, we have estimated the contribution from a correlated two-electron
process, in which a $1s$ electron
is excited and the $2s$ electron is emitted due to shake-off. We have calculated the
($E_e$-differential) shake probability by projecting the Li($2s$) state onto continuum
states of the single-center Hamiltonian
\begin{equation}
\hat h_{\rm shake} = -\frac{1}{2}\Delta_{r_t} - \frac{Z_T}{r_t} + 
          \frac{1}{r_t}[1-(1+Z_Tr_t)e^{-2Z_Tr_t}] .
\end{equation}
For $Z_T=3$ the total potential of this Hamiltonian corresponds 
to the sum of the Coulomb potential of the lithium
nucleus
and an electrostatic potential due to the presence of one hydrogenlike $1s$ electron,
i.e., the effects due to the presence of the excited electron and the repulsion 
between the $1s$ and $2s$ electrons are neglected. 
We have repeated the shake-off calculation with continuum states of the
Li$^+$ Hamiltonian used in the IEV calculation and have found very similar results
indicating that these simplifications are of minor importance.

\begin{figure}[ht]
\begin{center}
\includegraphics[width = \linewidth]{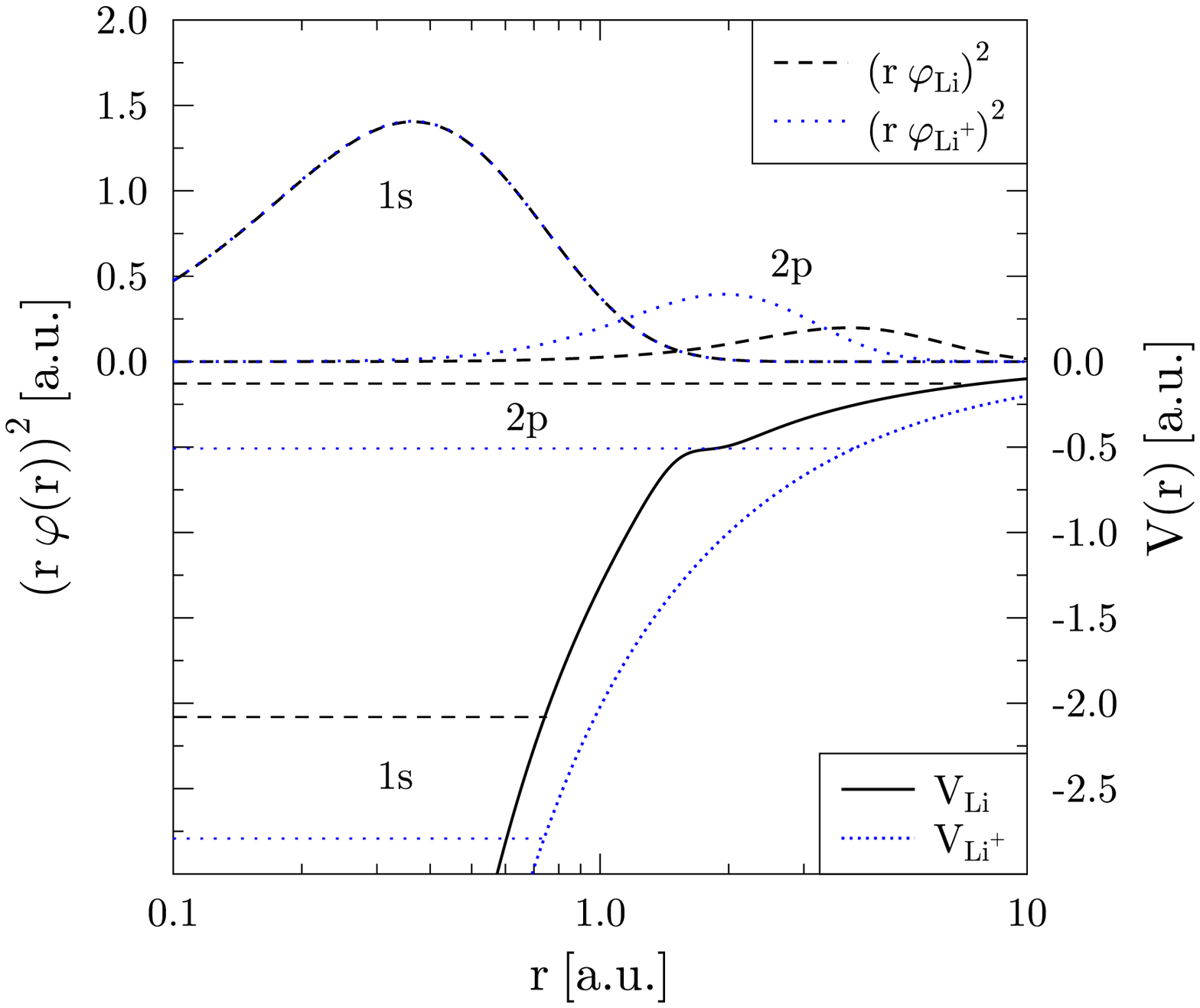}
\caption{(Color online) OPM potentials and $1s$ and $2p$ radial orbital densities for Li and Li$^+$.
The Li($1s$) and Li$^+$($1s$) densities sit on top of each other.
The horizontal lines indicate the binding energies of the $1s$ and $2p$ orbitals.
\label{fig:pot-orb}}
\end{center}
\end{figure}

Figure~\ref{fig:sdcs2} displays the results of these extensions.
The IEV SDCS~(\ref{eq:iev}) gives the strongest contribution. 
The correction due to the $n\rightarrow \infty$ extrapolation amounts to about 20\%
and is included in the cross section curve shown. The strongest partial channels 
are those corresponding to the dipole-allowed excitations into $2p$ ($\sim 44$\%) and $3p$ ($\sim 13$\%).
Interestingly, the IEV SDCS is much stronger
than its IEL counterpart for the process (\ref{eq:ei2}) shown in Fig.~\ref{fig:sdcs1}.
This mirrors the fact that the single-particle excitation probabilities 
for the lithium ion ($p_{1s+}^{\rm exc}$) 
are larger than those for the atom ($p_{1s}^{\rm exc}$), which can be understood by
inspecting the orbital densities of the single-particle states involved.
As shown in Fig.~\ref{fig:pot-orb} the Li$^+(1s)$ and
Li($1s$) densities are practically indistinguishable on a linear plot.
They only differ in their
asymptotic decays which are determined by the energy eigenvalues (indicated as horizontal
lines in Fig.~\ref{fig:pot-orb}).
By contrast, the
excited states are markedly different with those of Li$^+$ being more compact than those of Li 
and thus easier accessible for an initial $K$-shell electron. Figure~\ref{fig:pot-orb} shows that
in the case of the neutral lithium atom the $1s$ and $2p$ states have only very little
spatial overlap.
These observations favor the IEV picture, because the
ionic excited states are more realistic final states for EI than the neutral
atom states. 
Figure~\ref{fig:pot-orb} also displays the Li and Li$^+$ OPM potentials. 
The shoulder in the neutral atom potential is a reflection of the shell structure.

Coming back to Fig.~\ref{fig:sdcs2}, we observe that the $1s$ ionization with $2s$ excitation SDCS 
corresponding to Eq.~(\ref{eq:ei1}) 
increases by a little less than 20\% 
when the contribution from excitations to $n\ge 5$ is included,
and becomes 
very similar to the exclusive $1s$ ionization curve (which is repeated in Fig.~\ref{fig:sdcs2}).
Again the $2p$ excitation channel is the strongest one, contributing almost half of the cross 
section.
The shake-off SDCS, obtained by
replacing $p_{2s}^{\rm ion}$ in Eq.~(\ref{eq:ei2}) by the shake probability,
applying the $n\rightarrow \infty$ correction to $p_{1s}^{\rm exc}$ and integrating over 
the impact parameter, falls off steeply with increasing
electron energy, but it does give a non-negligible contribution 
at low $E_e$. When added to the three other contributions shown in Fig.~\ref{fig:sdcs2}
we obtain the thick full curve. The agreement with the 
measurements is not perfect, but it is very good down to $E_e\approx 2$~eV. 

\section{Summary}
\label{sec:summary}
In this work, we have used the independent-electron model to
analyze the contributions from single ionization with and without
additional target excitation to the recent MOTReMi SDCS measurements for 1.5 MeV/amu O$^{8+}$-Li 
collisions.
Our results, obtained by combining 
TC-BGM and CDW-EIS single-particle probabilities, largely confirm what was speculated
in Ref.~\cite{fischer12}: Two-electron excitation-ionization processes occur
and must be taken into account when comparing with the measurements for the $1s$ channel.
However, the independent-electron model is not sufficient to explain the data.
Rather, we had to replace it by an independent-event calculation for one of the
EI processes, i.e., we had to take relaxation effects into account, and we also
had to include a shake-off process to obtain good agreement with the measurements.
Accordingly, we have to conclude that electron correlation effects 
play some role in this collision system. 

It would be of interest to extend the experimental and theoretical studies by varying
the collision parameters projectile charge and velocity.
This would provide additional insight into the limitations of the independent-electron
model and the role of relaxation and shake-off processes.
Ultimately, the theoretical challenge consists in carrying out a full three-electron
calculation for this collision system. 

\begin{acknowledgments}
This work was supported by the Natural Sciences and Engineering Research Council of Canada (NSERC)
and by the Hungarian Scientific Research Fund (OTKA Grant No. K~109440).
We thank Eberhard Engel for making his atomic structure calculations available to us.
\end{acknowledgments}

\appendix*
\section{}
According to Eqs.~(\ref{eq:parthole}) and (\ref{eq:2svac}) the $2s$ vacancy production 
$P_{2s}^{\, \rm vac}$ is given as
\begin{equation}
  P_{2s}^{\, \rm vac} =  P_{1s\uparrow 1s\downarrow} - 
                  \sum_{f_k\in T} P_{1s\uparrow 1s\downarrow f_k\uparrow} .
\label{eq:p2svac-app1}
\end{equation}
The inclusive probability  $P_{1s\uparrow 1s\downarrow}$ is the determinant of the $2\times 2$ 
density matrix corresponding to the configuration $(1s\uparrow 1s\downarrow)$:
\begin{equation}
    P_{1s\uparrow 1s\downarrow} = 
        \left| \begin{array}{cc}
        \bra{1s\uparrow}\hat\gamma^1(t_f)\ket{1s\uparrow}  & \bra{1s\uparrow}\hat\gamma^1(t_f)\ket{1s\downarrow}  \\
        \bra{1s\downarrow}\hat\gamma^1(t_f)\ket{1s\uparrow}  & \bra{1s\downarrow}\hat\gamma^1(t_f)\ket{1s\downarrow} 
       \end{array}\right| ,
\label{eq:pincl1s2-1}
\end{equation}
where [cf. Eq.~(\ref{eq:gamma1})]
\begin{equation}
        \hat\gamma^1(t_f) = \ket{\psi_{1s\uparrow}(t_f)}\bra{\psi_{1s\uparrow}(t_f)} +
                            \ket{\psi_{1s\downarrow}(t_f)}\bra{\psi_{1s\downarrow}(t_f)} +
                             \ket{\psi_{2s\uparrow}(t_f)}\bra{\psi_{2s\uparrow}(t_f)} .
\label{eq:gamma1-exp}
\end{equation}
Inserting (\ref{eq:gamma1-exp}) into (\ref{eq:pincl1s2-1}), exploiting spin orthogonality 
and omitting the time argument for convenience we obtain
\begin{equation}
    P_{1s\uparrow 1s\downarrow} = 
        \left| \begin{array}{cc}
        |\braket{1s}{\psi_{1s}}|^2 + |\braket{1s}{\psi_{2s}}|^2   &  0 \\
         0  &  |\braket{1s}{\psi_{1s}}|^2
       \end{array}\right|  
    =  (p_{\rm 1s}^{\rm elast})^2 + p_{\rm 1s}^{\rm elast} p_{2s\rightarrow 1s} .
\label{eq:pincl1s2-2}
\end{equation}
In the last step we have used the definitions (\ref{eq:p1s-fk}), (\ref{eq:p2s-fk}),
and (\ref{eq:p1selast}).

The exclusive probability  $P_{1s\uparrow 1s\downarrow f_k\uparrow}$ is the determinant of the $3\times 3$ 
density matrix corresponding to the configuration $(1s\uparrow 1s\downarrow f_k\uparrow)$.
Using similar arguments as above we can write
\begin{eqnarray}
    P_{1s\uparrow 1s\downarrow f_k\uparrow} &=&
        \left| \begin{array}{ccc}
        |\braket{1s}{\psi_{1s}}|^2 + |\braket{1s}{\psi_{2s}}|^2   &  0 & \braket{1s}{\psi_{1s}} \braket{\psi_{1s}}{f_k} + \braket{1s}{\psi_{2s}} \braket{\psi_{2s}}{f_k} \\
         0  &  |\braket{1s}{\psi_{1s}}|^2  &  0  \\
        \braket{f_k}{\psi_{1s}} \braket{\psi_{1s}}{1s} + \braket{f_k}{\psi_{2s}} \braket{\psi_{2s}}{1s} & 0 & |\braket{f_k}{\psi_{1s}}|^2 + |\braket{f_k}{\psi_{2s}}|^2 
       \end{array}\right| \nonumber \\
    &=&
      (p_{\rm 1s}^{\rm elast})^2 p_{2s\rightarrow f_k}  +  p_{\rm 1s}^{\rm elast}[p_{1s\rightarrow f_k} p_{2s\rightarrow 1s}
       - (\braket{1s}{\psi_{1s}}\braket{\psi_{1s}}{f_k}
          \braket{f_k}{\psi_{2s}}\braket{\psi_{2s}}{1s} + c.c.) ] , 
\label{eq:pexcl1s2fk-1}
\end{eqnarray}
where $c.c.$ denotes the complex conjugate. 
One easily infers from 
the last expression of Eq.~(\ref{eq:pexcl1s2fk-1}) that $P_{1s\uparrow 1s\downarrow f_k\uparrow}=0$
for $\ket{f_k}=\ket{1s}$ as is required by the Pauli principle.

Inserting (\ref{eq:pincl1s2-2}) and (\ref{eq:pexcl1s2fk-1}) in (\ref{eq:p2svac-app1}) we obtain
\begin{eqnarray}
P_{2s}^{\, \rm vac} &=& (p_{\rm 1s}^{\rm elast})^2 \left(1-\sum_{f_k\in T}p_{2s\rightarrow f_k} \right)
          + p_{\rm 1s}^{\rm elast} p_{2s\rightarrow 1s} \left(1-\sum_{f_k\in T}p_{1s\rightarrow f_k} \right) 
\nonumber \\
  && \mbox{} +  p_{\rm 1s}^{\rm elast} \sum_{f_k\in T} (\braket{1s}{\psi_{1s}}\braket{\psi_{1s}}{f_k}
          \braket{f_k}{\psi_{2s}}\braket{\psi_{2s}}{1s} + c.c.)  .
\end{eqnarray}
In the last step
we use the definitions for single-particle ionization (\ref{eq:p1sion}) and
(\ref{eq:p2sion}) and remember that capture is negligible to obtain Eq.~(\ref{eq:2svac-result})
with the abbreviations (\ref{eq:2sexcl}) and (\ref{eq:2sex}) and the correction term
\begin{equation}
    \Delta P_{2s}^{\rm anti} = 
      p_{\rm 1s}^{\rm elast} \sum_{f_k\in T} (\braket{1s}{\psi_{1s}}\braket{\psi_{1s}}{f_k}
          \braket{f_k}{\psi_{2s}}\braket{\psi_{2s}}{1s} + c.c.) .
\label{eq:deltap2s}
\end{equation}

Equation~(\ref{eq:1svac-result}) for the $1s$ vacancy production $P_{1s}^{\, \rm vac}$ 
can be derived using similar arguments. The calculation is
straightforward but lengthy. One finds for the correction term: 
\begin{eqnarray}
    \Delta P_{1s}^{\rm anti} &=& 
      - p_{\rm 1s}^{\rm ion} \sum_{f_k\in T, f_k\neq 1s} 
          (\braket{1s}{\psi_{1s}}\braket{\psi_{1s}}{f_k}
          \braket{f_k}{\psi_{2s}}\braket{\psi_{2s}}{1s} + c.c.) 
\nonumber \\
      && \mbox{} +  p_{\rm 1s}^{\rm exc} \sum_{f_k\in T} 
           (\braket{1s}{\psi_{1s}}\braket{\psi_{1s}}{f_k}
          \braket{f_k}{\psi_{2s}}\braket{\psi_{2s}}{1s} + c.c.) 
\nonumber \\
      && \mbox{} +  p_{\rm 1s}^{\rm elast} \sum_{f_k\in T}\sum_{f_l\in T, f_l\neq 1s}
          (\braket{f_k}{\psi_{1s}}\braket{\psi_{1s}}{f_l}
          \braket{f_l}{\psi_{2s}}\braket{\psi_{2s}}{f_k} + c.c.) .
\label{eq:deltap1s}
\end{eqnarray}

\bibliography{lithium-iev}

\end{document}